\documentstyle[12pt,psfig]{l-aa}
\def\msun{M_{\odot}}
\def\ergsec{\hbox{erg s$^{-1}$}}

\def\degmark{^\circ}
\def \rsun {\ifmmode$R$_{\odot}\else R$_{\odot}$\fi}
\def \nh {N${\rm _H}$}
\def \hcm {\hbox {\ifmmode $ atoms cm$^{-2}\else atoms cm$^{-2}$\fi}}
\def \src {CAL\,83}
\def\approxgt{\mathrel{\hbox{\rlap{\lower.55ex \hbox {$\sim$}}
        \kern-.3em \raise.4ex \hbox{$>$}}}}
\def\approxlt{\mathrel{\hbox{\rlap{\lower.55ex \hbox {$\sim$}}
        \kern-.3em \raise.4ex \hbox{$<$}}}}
\newcommand {\rosat} {{ROSAT}}
\newcommand {\einstein} {{Einstein}}

\newcommand {\asca} {{ASCA}}
\newcommand {\sax} {BeppoSAX}

\newcommand {\Msun} {{ M$_{\odot}$}}

\newcommand {\rchisq} {$\chi_{{\rm \nu}} ^{2}$}

\begin{document}

\thesaurus{ (02.01.2; 08.02.1; 08.09.2; 08.23.1; 13.25.5)}

\title{A \sax\ LECS observation of the super-soft X-ray source \src }

\author{A.N. Parmar\inst{1} \and P. Kahabka\inst{2} \and H.W.
Hartmann\inst{3} \and J. Heise\inst{3} \and B.G. Taylor\inst{1}}

\institute{
Astrophysics Division, Space Science Department of ESA, 
ESTEC, P.O. Box 299, NL-2200 AG Noordwijk, The Netherlands
\and 
Astronomical Institute and Center for High Energy Astrophysics,
University of Amsterdam, Kruislaan 403, NL-1098 SJ Amsterdam, The
Netherlands
\and 
SRON Laboratory for Space Research, Sorbonnelaan 2, 
NL-3584 CA Utrecht, The Netherlands}

\date{Received ; accepted}
\offprints{A.N. Parmar: aparmar@astro.estec .esa.nl}
\maketitle

\begin{abstract}

We report on a \sax\ Low-Energy Concentrator Spectrometer (LECS)
observation of the super-soft source (SSS) \src. The X-ray emission
in SSS is believed to arise from nuclear burning of accreted material
on the surface of a white dwarf (WD).
The LECS spectrum of \src\ can be well fit by both absorbed
blackbody and WD atmosphere models. If the
absorption is constrained to be equal to the value derived from
{\it Hubble Space Telescope} measurements,
then the best-fit blackbody temperature is $46.4 \pm 1.4$~eV while
a Non Local Thermal Equilibrium (NLTE) WD 
atmosphere model gives a lower temperature of $32.6 \pm 0.7$~eV. 
In contrast to CAL\,87, there are no strong absorption edges visible
in the X-ray spectrum with a 68\% confidence upper limit of 2.3
to the optical depth of a C~{\sc vi} edge at 0.49~keV predicted by
WD atmosphere models.
The luminosity and radius derived from the NLTE
fit are consistent with the values predicted for stable nuclear burning
on the surface of a $\sim$0.9--1.0$\msun$ WD.

\end{abstract}


\section{Introduction}
\label{sec:introduction}

The \einstein\ observatory performed a survey of the Large Magellanic Cloud
(LMC) in which two sources with unusually soft spectra, \src\ and CAL\,87 were
detected (Long et al. 1981). These sources emit little or no radiation
at energies $\approxgt$1~keV and became known as ``super-soft'' sources
(SSS). Subsequent
\rosat\ and \asca\ observations have revealed approximately 30 similar sources
located in the Galaxy, the Magellanic Clouds, a globular cluster and M31
(see Kahabka \& Tr\"umper 1996; van Teeseling 1997 for recent reviews). 
\src\ is often regarded as the prototypical SSS. It is identified with
a m$_{\rm _V} = 17$ variable blue stellar object which shows strong
H${_\beta}$ and He {\sc ii} $\lambda$4686 emission and 
an orbital period of 1.04 days (Cowley et al. 1984; Pakull et al. 1985;
Smale et al. 1988; Crampton et al. 1987). \src\ is surrounded by a weak
photoionized nebula (Pakull \& Angebault 1986).
A blackbody fit to the \rosat\ Position Sensitive Proportional Counter
(PSPC) spectrum indicates a temperature of $<$40~eV, an absorbing column 
compatible with the LMC value, and a luminosity of $>$$10^{39}$~\ergsec\
(Greiner et al. 1991). A {\it Hubble Space Telescope} (HST) spectrum of \src\
reveals numerous interstellar lines as well as emission features due
to O. The depth of a broad Ly$\alpha$ profile 
corresponds to a neutral hydrogen column of $(6.5 \pm 1.0) \times
10^{20}$~\hcm\ (G\"ansicke et al. 1997). 

Van den Heuvel et al. (1992) proposed that SSS are systems undergoing
steady nuclear burning of hydrogen accreted onto the surface of a white 
dwarf (WD) with masses in the range 0.7--1.2\Msun. The mass transfer from
a main-sequence or sub-giant companions is unstable on a thermal 
time scale and for a narrow range of accretion rates, steady nuclear
burning can take place. Evolutionary scenarios for such systems are
discussed in Rappaport et al. (1994). It is unlikely that SSS
compose a homogeneous class and one way of probing the nature
of individual sources is by searching for the characteristic spectral
signatures of nuclear burning on a WD. This burning takes
place deep within the WD atmosphere at a large energy
dependent optical depth.
Photoelectric absorption by highly ionized metals in the atmosphere can
produce edges in the X-ray spectrum.
\sax\ observations of CAL\,87 have revealed the presence of a deep
O~{\sc viii} edge at an energy of 0.871~keV (Parmar et al. 1997a).
Such edges are a common feature of WD atmosphere models that 
assume Local Thermodynamic Equilibrium
(LTE) such as that of Heise et al. (1994) and also of 
non-LTE (NLTE) models such as that of Hartmann \& Heise (1997a). 
The CAL\,87 spectral fits support the view that the X-ray emission from
at least one SSS results from nuclear burning in the atmosphere of a WD.
As part of a systematic study of SSS spectra undertaken with \sax,
we report results from an observation of \src.  

\begin{figure*}
  \centerline{\hbox{
         \psfig{figure=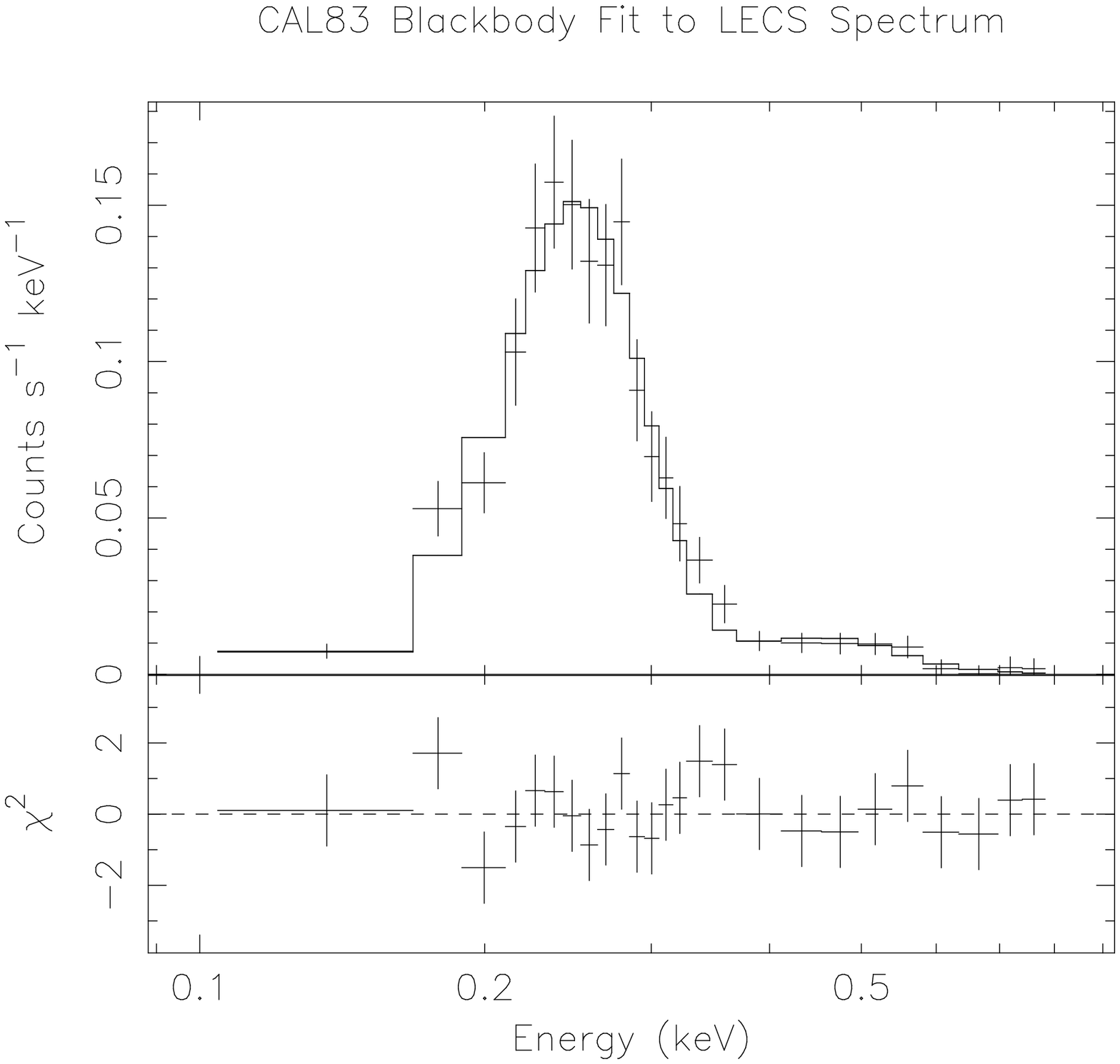,height=8.5cm,angle=-90}
         \hspace{-2.5cm}
         \psfig{figure=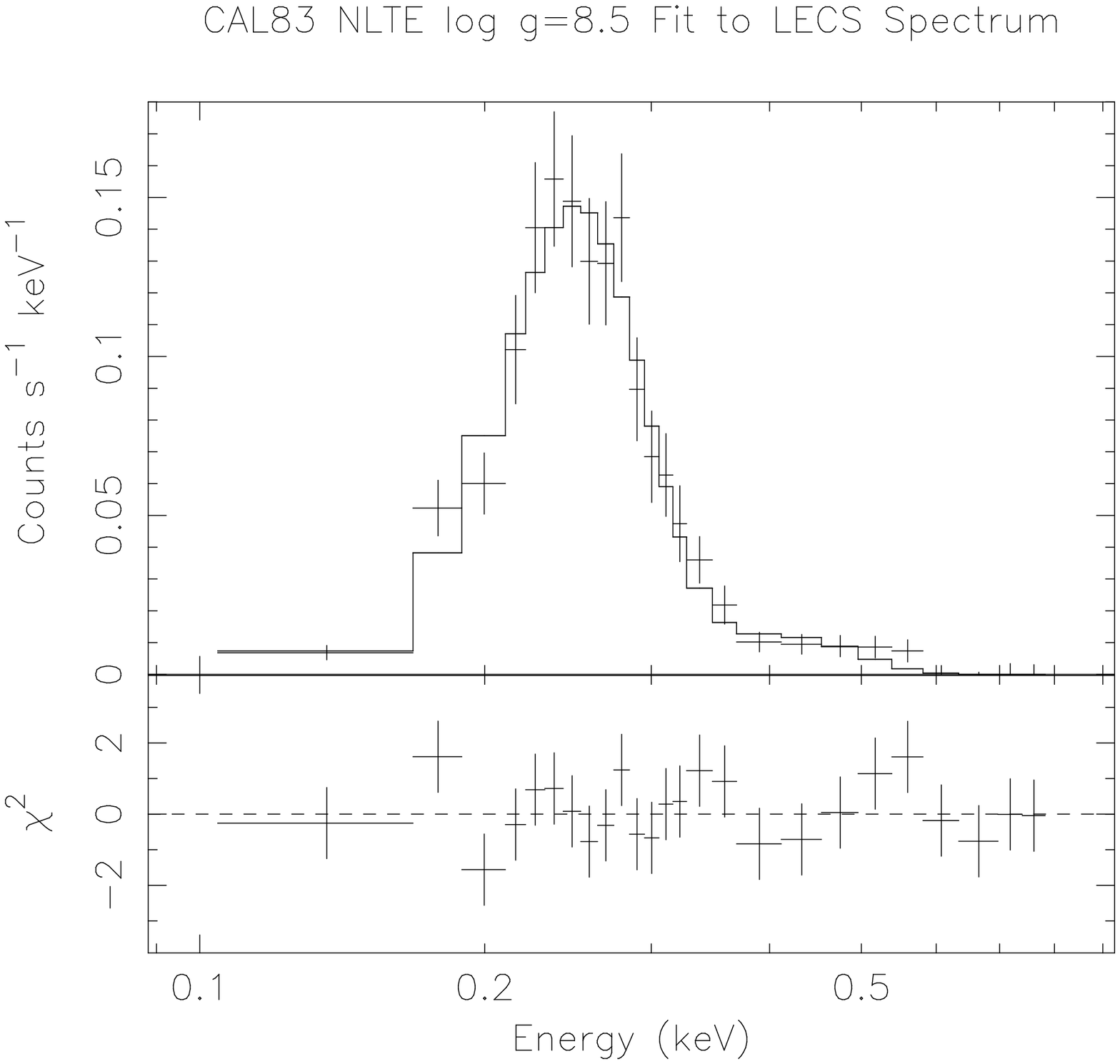,height=8.5cm,angle=-90}
}}
  \caption[]{Best-fit absorbed blackbody (left-panel) and NLTE 
(right panel) model fits to the LECS \src\ spectrum}
  \label{fig:fits}
\end{figure*}

\section{Observations}
\label{sec:observations}

The scientific payload of the \sax\ X-ray Astronomy Satellite 
(Boella et al. 1997) comprises 
four coaligned Narrow Field Instruments, or NFI, including the 
LECS. This is an 
imaging gas scintillation proportional counter sensitive in 
the energy range 0.1--10.0~keV with a circular field of view of 37$'$ 
diameter (Parmar et al. 1997b). The usual background counting rate is
$9.7 \times 10^{-5}$~arcmin$^{-2}$~s$^{-1}$ in the energy range 
0.1--10.0~keV. The LECS energy
resolution is a factor $\sim$2.4 better than that of the \rosat\ PSPC,
while the effective area is between a factor $\sim$6 and 2 lower at 0.28 and
1.5~keV, respectively.   
\src\ was observed by the LECS 
between 1997 March 07 04:04 and March 08 15:58~UTC. 
Good data were selected from intervals when the minimum elevation angle
above the Earth's limb was $>$4$\degmark$ and when the instrument parameters
were nominal using the SAXLEDAS 1.7.0 data analysis package.  
Since the LECS was only operated 
during satellite night-time, this gave a total on-source exposure of
36~ks. 

Examination of the LECS image shows a source at a position consistent
with (16$''$ distant) that of \src.
A spectrum was extracted centered on the source centroid using a
radius of 8$'$. This radius was chosen to include 95\% of the 0.28~keV
photons.
The spectrum was rebinned to have $>$20
counts in each bin to allow the use of the $\chi^2$ statistic. The XSPEC
9.01 package (Arnaud 1996)
was used for spectral analysis together with the response matrix 
from the 1997 September release.
During the observation the particle induced background was a factor
of $\sim$2 higher than usually encountered. For this reason
the background spectrum was extracted from the image itself using
an annulus centered on \src\ with inner and outer radii of 9$'$ and 20$'$, 
respectively. A small correction for telescope vignetting was applied.
The \src\ count rate above background is
$0.01965 \pm 0.00084$~s$^{-1}$. Examination of the extracted spectrum
shows that the source is only detected in a narrow energy range (see
Fig.~\ref{fig:fits}) and only the 25 rebinned channels 
corresponding to energies between 0.1 and 0.8~keV were used for spectral 
fitting. 

\subsection {Blackbody spectral fits}
\label{subsect:bb_fits}
\begin{figure*}
  \centerline{\psfig{figure=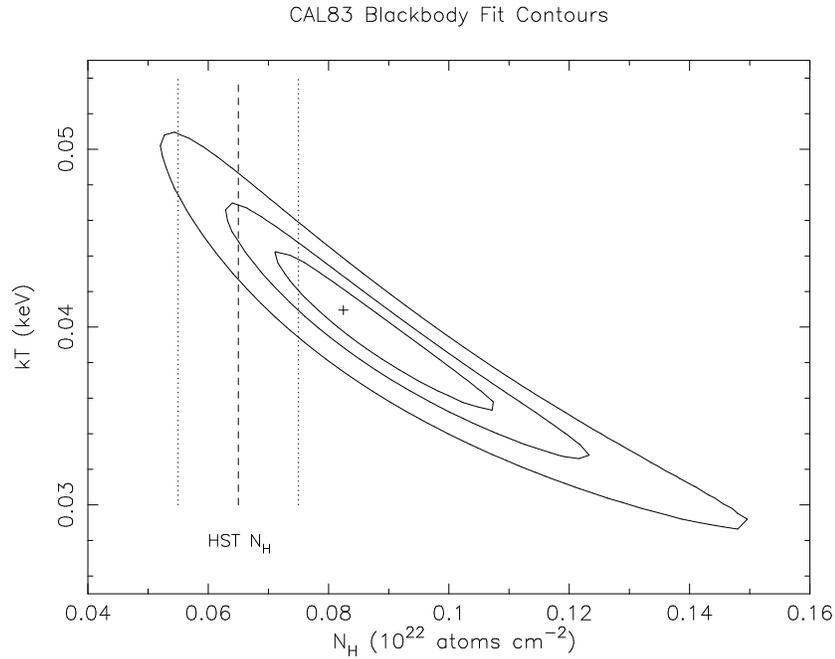,width=12.0cm,angle=-90}}
  \caption[]{$\chi ^2$ contours of blackbody fits to the LECS \src\ 
  spectrum. The contours correspond to 68\%, 90\%, and 99\% confidence
  levels. The best-fit is indicated with a cross. The value of N$_{\rm H}$ 
  derived from HST measurements by G\"ansicke et al. (1997) is indicated 
  along with the $\pm$1$\sigma$ uncertainties as dotted lines}
  \label{fig:contours}
\end{figure*}

In order to compare the LECS spectrum with those obtained from previous
observations, an absorbed blackbody spectral model was first fit to the data
(Fig.~\ref{fig:fits}).
The photoelectric absorption coefficients of Morisson \& McCammon (1983) 
together with the solar abundances of Anders \& Grevesse (1989) were used.
An acceptable fit is obtained with \rchisq\ of 0.71 for 22 degrees
of freedom (dof). The best-fit parameters are given in Table~\ref{tab:bb_fits}
and the uncertainty contours are displayed in Fig.~\ref{fig:contours}.
A distance of 50~kpc is assumed in order to
derive the WD radius, R, and luminosity, L, and
all uncertainties are quoted at 68\% confidence corresponding to 
${\rm \chi^2 _{min} + 1.0}$.

The uncertainties on the temperature, luminosity and radius can be
further constrained if the absorption, \nh, is fixed at the best-fit value
derived from the HST measurements by G\"ansicke et al. (1997) of
$6.5 \times 10^{20}$~\hcm. The best-fit temperature is now somewhat
higher ($46.4 \pm 1.4$~eV compared to $41 \pm 6$~eV) and much
more tightly constrained leading to improved estimates of the emission
region radius, luminosity and temperature of 
$(8.26 \pm 0.30)\times 10^8$~cm, $(2.49 \pm 0.19)\times 10^{37}$~\ergsec\ and 
$46.4 \pm 1.4$~eV, respectively. The slightly higher best-fit \nh\ value
may indicate the presence of absorption local to the X-ray source.

\begin{table}
\caption[]{\src\ blackbody spectral fit parameters}
\begin{flushleft}
\begin{tabular}{lll}
\hline\noalign{\smallskip}
Model& \multicolumn{2}{c}{Blackbody} \\
\noalign {\smallskip}
\hline\noalign {\smallskip}
      T (eV)                     & $41\pm 6$ & $46.4 \pm 1.4$ \\
      \nh\ ($10^{20}$ cm$^{-2}$) & $8.2 \pm ^{1.9}_{1.1}$ 
                                 & 6.5$^{\rm a}$ \\ 
      R ($10^9$~cm)              & $1.33 \pm ^{1.41} _{0.36}$ 
                                 & $0.826 \pm 0.030$ \\
      L (10$^{37}$ erg s$^{-1}$)& $6.4 \pm ^{21.1} _{3.0}$ 
                                 & $2.49 \pm 0.19$\\
      \rchisq\                   & 0.71 & 0.76 \\
      dof                        & 22 & 23 \\
\noalign {\smallskip}
\hline
\multicolumn{3}{l}{\footnotesize $^a$Fixed at HST value of
G\"ansicke et al. (1997)}\\
\end{tabular}
\end{flushleft}
\label{tab:bb_fits}
\end{table}

\subsection{WD model atmosphere spectral fits}
\label{subsect:model_fits}

Parmar et al. (1997a) demonstrate that the X-ray spectrum of CAL\,87
is better described by hot WD atmosphere models undergoing hydrogen
burning than with a simple blackbody.
In order to see if this is true for \src, LTE and NLTE WD atmosphere
models were fit to the LECS spectrum.
In both the LTE and NLTE models plan parallel
geometry and hydrostatic equilibrium are assumed and model spectra
are calculated for different surface gravities, elemental abundances and
temperatures. Bound-free opacities for the ground states of 
H, He, C, N, O, Ne, Na, Mg, Si, S, Ar, Ca and Fe have been included 
(Hartmann \& Heise 1997b). The effects of line opacities and 
line-blanketing are not included. This will 
change the derived atmospheric temperature structure 
and shift the ionization balance (Hubeny \& Lanz 1995a,b).
As a result the depths of the edges will be altered with respect to the
unblanketed spectra and consequently, due to the constraint of 
radiative equilibrium, the overall shape of the continuum will change.
Details of line-blanketing effects in hot high gravity NLTE white
dwarf model atmospheres will be reported in Hartmann \& Heise (1998).
A surface gravity, $g$, of $10^9$~cm~s$^{-2}$, appropriate 
to a $\sim$1.0$\msun$ WD is assumed and the results
are given in Table~\ref{tab:lte_fits}.
Initial NLTE fits were performed using models cf. Hartmann
\& Heise (1997b) for surface gravities of
log $g$ = 8.0, 8.5 and 9.0, assuming an
LMC abundance of 0.25 times the solar value and allowing \nh\ to 
vary. 
The fit results are  
insensitive to abundance (see Sect.~\ref{sect:discussion})
and adopting solar abundance gives almost
identical values. 
Examination of Tables~\ref{tab:lte_fits} and ~\ref{tab:nlte_fits} 
reveals that none of the WD atmosphere models give as good a fit
as a simple blackbody. In addition, as $g$ increases the best-fit 
values of T, \nh, R and L derived from the NLTE fits all increase. 
If \nh\ is constrained to
have the HST value of G\"ansicke et al. (1997) of $6.5 \times 10^{20}$~\hcm,
then the best fit is obtained with $\log g = 8.45$. 
This gives best-fit values for T, R, and L of $32.6 \pm 0.7$~eV,
$(1.25 \pm ^{0.08}_{0.05}) \times 10^9$~cm, and 
$(2.30 \pm ^{0.10} _{0.05}) \times 10^{37}$~\ergsec,
respectively.

In order to set limits on the luminosity of any additional
spectral component, a thermal bremsstrahlung model
was added to the NLTE WD atmosphere model. The value of \nh\ was 
constrained as before. The
\rchisq\ did not improve significantly and the 68\% confidence upper-limit
to the best-fit 0.4~keV bremsstrahlung component is 
$5.3 \times 10^{35}$~\ergsec. 

\begin{table}
\caption[]{\src\ LTE (Heise et al. 1994) spectral fit parameters}
\begin{tabular}{ll}
\hline\noalign{\smallskip}
Model & LTE ($\log g = 9.0$) \\
\noalign {\smallskip}
\hline\noalign {\smallskip}
      T (eV)                     & $44.0 \pm ^{6.3} _{0.6}$ \\ 
      \nh\ ($10^{20}$ cm$^{-2}$) & $9.5 \pm ^{0.3} _{5.2}$ \\
      R ($10^9$~cm)              & $1.05 \pm ^{0.08} _{0.76}$\\
      L  (10$^{37}$ erg s$^{-1}$)& $5.4 \pm ^{0.6} _{4.7}$ \\
      \rchisq\                   & 1.02 \\
      dof                        & 22 \\
\noalign {\smallskip}
\hline
\end{tabular}
\label{tab:lte_fits}
\end{table}

\begin{table}
\caption[]{\src\ NLTE (Hartmann \& Heise 1997b) spectral fit parameters}
\begin{flushleft}
\begin{tabular}{lll}
\hline\noalign{\smallskip}
Model & NLTE ($\log g = 8.0$) & NLTE ($\log g = 8.45$) \\
\noalign {\smallskip}
\hline\noalign {\smallskip}
      T (eV)                     & $31.7 \pm ^{0.5} _{3.8}$ 
                                 & $32.6 \pm 0.7$ \\
      \nh\ ($10^{20}$ cm$^{-2}$) & $5.4 \pm ^{4.0} _{2.2}$ 
                                 & 6.5$^a$\\ 
      R ($10^9$~cm)              & $1.1 \pm ^{3.2} _{0.7}$ 
                                 & $1.25 \pm ^{0.08} _{0.05}$ \\
      L  (10$^{37}$ erg s$^{-1}$)& $1.6 \pm ^{7.9} _{1.0}$ 
                                 & $2.30 \pm ^{0.10} _{0.05}$\\
      \rchisq\                   & 0.92 & 0.76 \\
      dof                        & 22 & 23 \\
\noalign {\smallskip}
\hline
\noalign {\smallskip}
Model  & NLTE ($\log g = 8.5$) & NLTE ($\log g = 9.0$) \\
\noalign {\smallskip}
\hline\noalign {\smallskip}
      T (eV)                     & $32.4 \pm 1.2$ 
                                 & $34.15 \pm ^{2.8} _{0.35}$ \\ 
      \nh\ ($10^{20}$ cm$^{-2}$) & $6.8 \pm ^{1.4} _{1.0}$  
                                 & $7.8 \pm ^{0.5} _{2.3}$ \\
      R ($10^9$~cm)              & $1.35 \pm ^{0.6} _{0.3}$ 
                                 & $1.4 \pm ^{2.7} _{0.7}$ \\   
      L  (10$^{37}$ erg s$^{-1}$)& $2.6 \pm ^{2.0} _{0.8}$ 
                                 & $3.4 \pm ^{0.7} _{2.1}$ \\ 
      \rchisq\                   & 0.79 & 0.81 \\
      dof                        & 22  & 22 \\
\noalign {\smallskip}
\hline
\multicolumn{3}{l}{\footnotesize $^a$Fixed at HST value of
G\"ansicke et al. (1997)}\\
\end{tabular}
\end{flushleft}
\label{tab:nlte_fits}
\end{table}

\section {Discussion}
\label{sect:discussion}

The LECS spectrum of \src\ is well fit by all three types of 
trial model and a considerably 
longer exposure is required in order to meaningfully discriminate between 
these models based on fit quality alone.  
There are however significant differences in the best-fit 
parameters derived using 
the different models (see Tables~\ref{tab:bb_fits}, \ref{tab:lte_fits},
and \ref{tab:nlte_fits}) with the best-fit blackbody and LTE 
temperatures higher than with the NLTE fits, and correspondingly smaller
emission region radii.
If \nh\ is constrained to be equal to the HST value, then both the
blackbody and NLTE fits give similar sub-Eddington
luminosities of $\sim$$2.5 \times 10^{37}$~\ergsec,
but the NLTE fit gives a significantly lower temperature of 
$32.6 \pm 0.7$~eV,
compared to $46.4 \pm 1.4$~eV for the blackbody fit. 

The good fits to all three types of spectral model is in contrast to 
the LECS spectrum
of CAL\,87, which is better fit by both LTE and NLTE 
WD atmosphere models than a blackbody. This is mainly due to the
presence of a deep (a best-fit optical depth $>$13) O~{\sc viii} 
absorption edge at an
energy of 0.871~keV (Parmar et al. 1997a).
The CAL\,87 results supports the view that the X-ray 
emission from SSS arises from nuclear burning on the surface of a WD, and
the same result would be expected from \src.
However, at the temperatures and gravities appropriate to the \src\
spectral fits, all the abundant ions that have absorption edges in the
energy range 0.2--0.6~keV are completely ionized, with the exception of
C~{\sc vi} at 0.49~keV. This depth of this edge is not predicted to be
as deep as that of the O~{\sc viii} edge seen from CAL\,87.  
The 68\% confidence upper limit optical 
depth to an edge at an energy of 0.49~keV is 2.3.
This probably explains why smooth functions such as blackbodies provide
good fits to the \src\ spectrum.
 
\begin{figure}
  \centerline{\psfig{figure=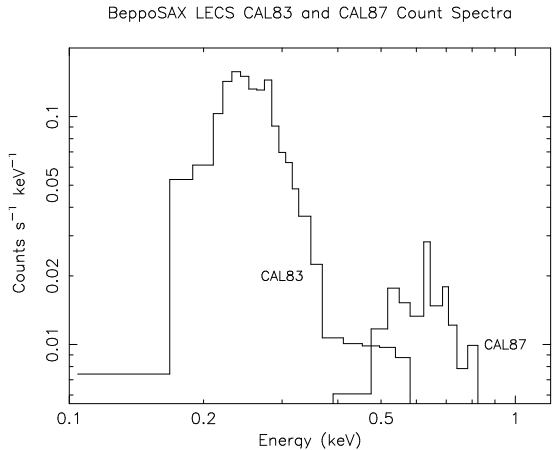,width=8.0cm,angle=-90}}
  \caption[]{Observed LECS count spectra for CAL\,87 (taken from
   Parmar et al. (1997a) and \src} 
  \label{fig:comparison}
\end{figure}

Comparison of the two panels in Fig.~\ref{fig:fits} reveals that
most of the additional contribution to $\chi ^2$ for the NLTE models 
is at energies $\approxgt$0.4~keV where the
blackbody provides a better fit. 
One possibility is that the 
model atmosphere data
are less appropriate at the temperatures used for modeling \src\ (30--50~eV),
rather than CAL\,87 (55--75~eV).
Some idea of the very different
LECS count spectra for these two sources can be seen in 
Fig.~\ref{fig:comparison}, although much of the difference is 
due to the $\sim$10 times lower absorption to \src\ 
compared to CAL\,87 ($\sim$$10^{21}$ and $\sim$$10^{22}$~\hcm, respectively).
Alternatively, we cannot exclude the possibility of a hard
component from \src\ with a luminosity of $\approxlt$$5\times 10^{35}$~\ergsec\
(see Sect.~\ref{subsect:model_fits}), which may affect the fits in the
$\approxgt$0.4~keV energy range.

The WD mass can be constrained assuming that \src\ is on the
stability line for steady nuclear burning (see Iben 1982, Fig.~2). 
The best-fit NLTE temperature,
when \nh\ is constrained to be equal to the HST value, of 
$32.6 \pm 0.7$~eV
corresponds to a WD of mass $\sim$0.9--1.0$\msun$ with a 
luminosity of $\sim$$3 \times 10^{37}$~\ergsec\ 
(see also Iben \& Tutukov 1996). This is very similar to the
measured luminosity of $(2.55 \pm 0.35) \times 10^{37}$~\ergsec.
This good agreement is in contrast to CAL\,87 where the luminosity
derived from the stability line is at least a factor 8 higher than
that observed (Parmar et al. 1997a). 
This implies that in the case of \src, the line of
sight to the emitting
region is not significantly obscured by e.g., an accretion disk.
In addition, the emission region radius of $(1.35 \pm 0.32) \times 10^9$~cm
is somewhat larger than the expected WD radius ($6.4 \times 10^8$~cm for
a 0.9$\msun$ WD; Bergeron et al. 1992).

In 1996 April \src\ underwent a temporary X-ray off-state (Kahabka 1997).
This has been modeled by either the cessation of steady nuclear
burning in the surface layers of a massive ($\approxgt$1.3$\msun$) WD (Alcock 
et al. 1997) or by the expansion of the WD atmosphere following an episode of
increased mass accretion rate on a massive ($>$1.2~$\msun$) WD (Kahabka 1997). 
Both these models support the view that the X-ray emission from \src\ 
originates from steady nuclear burning in the surface layers of a WD and
it is interesting to note that the WD mass derived above, assuming that the
system is on the stability line of Iben (1982), is lower than both the
above estimates.
The best-fit blackbody spectral parameters derived here,
approximately one year after the temporary off-state, are consistent
with those derived by Kahabka (1997) using the \rosat\ PSPC 
prior to the off-state. This
indicates that \src\ has returned to its pre off-state behavior.

In summary, the LECS spectrum of \src\ is consistent with
the assumption of a hot WD atmosphere heated by nuclear burning, but 
formally does not prove such an assumption.

\begin{acknowledgements}
We thank the staff of the \sax\ Science Data Center
for help with these observations. K. Ebisawa is thanked for
discussions.
PK is a Human Capital and Mobility Fellow.
The \sax\ satellite is a joint Italian and Dutch programme.
We thank the referee, M. Barstow, for helpful comments.
\end{acknowledgements}

\end{document}